\documentclass[global,referee]{svjour}
\usepackage{graphics}
\journalname{myjournal}
\begin{document}
\title{Polarisation dynamics of a birefringent Fabry-Perot cavity}
\author{Aldo Ejlli\inst{1} \and Federico Della Valle\inst{2,3} 
\and Guido Zavattini\inst{1}}


%
%
\institute{INFN, Sez. di Ferrara and Dip. di Fisica e Scienze della Terra, Universit\`a di Ferrara, via Saragat 1, Edificio C, I-44122 Ferrara, Italy \label{addr1} \and INFN, Sez. di Trieste and Dip. di Fisica, Universit\`a di Trieste, via A. Valerio 2, I-34127 Trieste, Italy \label{addr2} 
\and e-mail: federico.dellavalle@ts.infn.it}
\date{Received: date / Revised version: date}
%
\maketitle
\begin{abstract}
Optical Fabry-Perot cavities always show a non-degeneracy of two orthogonal polarisation states. This is due to the unavoidable birefringence of dielectric mirrors whose effects are extremely important in Fabry-Perot based high-accuracy polarimeters. We have developed and present here a theory of the polarisation state dynamics in a birefringent Fabry-Perot resonator, and we validate it through measurements performed with the polarimeter of the PVLAS experiment. The measurements are performed while a laser is frequency-locked to the cavity, and provide values for the finesse of the cavity and for the phase difference between the two orthogonal polarisation components introduced by the combination of the two cavity mirrors (equivalent wave-plate). The theoretical formulas and the experimental data agree well showing that the consequences of the mirror birefringence must be taken into account in this and in any other similar experiment.
\end{abstract}

\section{Introduction}
\label{intro}
Today high finesse Fabry-Perot (FP) cavities are often used to increase the effective optical path length of light within a given region. One such application is in very sensitive polarimetry \cite{PVLAS2015,Q&A2007,Cadne2014,Fan2017}. Typically, to further increase the sensitivity of such an apparatus, the effect to be measured is modulated in time. Finesses have become so high that more and more often the modulated effect has frequency components close to or above the frequency cutoff of the cavity itself. For example, in ellipticity measurements, such Fabry-Perot cavities are treated as first order filters \cite{Uehara1995,Berceau2010}. A complication exists when the mirrors of the Fabry-Perot cavity are birefringent, in that not only does the ellipticity generate a rotation \cite{PVLAS2015,Zavattini2006} but the ellipticity and the rotation have different frequency responses, which depend on the cavity intrinsic birefringence.

In this paper we present an experimental method to make a complete characterisation of the polarisation dynamics of a Fabry-Perot cavity used for polarimetry. The study should be of interest for the wide community that employs Fabry-Perot cavities to pursue measurements of fundamental physics.

\section{Polarimetry with a birefringent cavity}
\label{sec:1}

\subsection{General method}

\begin{figure}[tbh]
\begin{center}
\resizebox{.9\textwidth}{!}{%
  \includegraphics{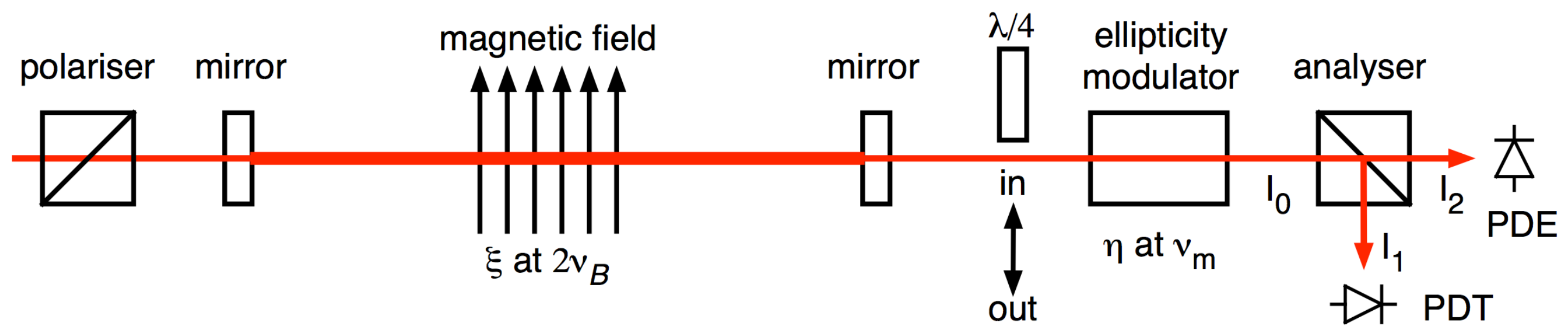}
}\end{center}
\caption{Principle scheme for magneto-optical polarimetry with a Fabry-Perot cavity. PDT: transmission photodiode that collects the intensity $I_1$ of the electric field component with polarisation equal to that of the input light; PDE: extinguished beam photodiode that collects the intensity $I_2$ of the orthogonal component.}
\label{fig:scheme}
\end{figure}

In previous papers \cite{PVLAS2015,Iacopini1979,Iacopini1981,BFRT1993,PVLAS1998,PVLAS2008,PVLAS2013,PVLAS2014}, the method employed by the PVLAS experiment for ultra-high-sensitive magneto-optical polarimetry based on a Fabry-Perot (FP) cavity was presented. The instrument is a Malus interferometer \cite{Vallet1999}, namely a Fabry-Perot cavity inserted between two crossed polarisers, with heterodyne detection \cite{Iacopini1979}. The principle scheme is shown in Fig.~\ref{fig:scheme}. 
The instrument measures the ellipticity and the rotation acquired by linearly polarised light as a consequence of a magnetic anisotropy of the complex index of refraction $\tilde{n}=n+i\kappa$ of the medium between the mirrors. The ellipticity is the ratio of the minor to the major axis of the ellipse described by the electric field vector. With respect to the axes of the ellipse, there is a phase difference of $\pi/2$ between the two orthogonal components of the electric field: if the component of the electric field along the major axis is real, the orthogonal component is a purely imaginary number. Differently, in rotations the two components remain in phase. We also remind the reader that, if all acquired ellipticities and rotations are small, then they add algebraically.

The most general element describing linear magnetic birefringence and dichroism can be expressed as a Jones matrix \cite{Jones1948} as
\[
\left(\begin{array}{cc}e^{\xi/2}&0\\0&e^{-\xi/2}\end{array}\right),
\]
where $\xi=i2\psi+2\theta$; an overall attenuation factor has been neglected. 
The ellipticity $\psi$ and the rotation $\theta$ acquired in a single passage through the magnetic region of length $L$ are given by
\[
\psi=\pi\frac{\Delta nL}{\lambda}\sin2\phi\equiv\psi_0\sin2\phi\qquad{\rm and}\qquad\theta=\pi\frac{\Delta\kappa L}{\lambda}\sin2\phi\equiv\theta_0\sin2\phi,
\]
where $\lambda$ is the light wavelength, $\Delta n=n_\parallel-n_\perp$ is the magnetic birefringence, $\Delta\kappa=\kappa_\parallel-\kappa_\perp$ is the magnetic dichroism and $\phi$ is the angle between the input polarisation direction (electric field of the light beam) and the external magnetic field. The subscripts $\parallel$ and $\perp$ refer to the direction of the external magnetic field. In the PVLAS experiment, rotating the dipole permanent magnets that generate the magnetic field modulates the magneto-optical effects; an ellipticity modulator is used for heterodyne detection. If a quarter-wave plate is properly inserted between the output mirror and the modulator, so as to transform rotations into ellipticities, rotation measurements are obtained. If the $\lambda/4$ plate is not inserted, ellipticity is measured. Indeed, the ellipticity of the modulator will only beat with other ellipticities. The Fabry-Perot lengthens the optical path within the interaction region thus amplifying both the ellipticity and the rotation by a factor $N=2{\cal F}/\pi$, where ${\cal F}$ is the finesse of the cavity given by \cite{BornWolf}
\begin{equation}
{\cal F}=\frac{\pi\sqrt{R}}{1-R},
\label{eq:finesse}
\end{equation}
with $R$ the reflectance of the cavity mirrors, assumed identical. Finesse values cas high as $7.7\times10^5$ can be obtained \cite{DellaValle2014OE}. We denote with
\[
\Psi=N\psi\equiv\Psi_0\sin2\phi\qquad{\rm and}\qquad\Theta=N\theta\equiv\Theta_0\sin2\phi\]
the amplified values of the ellipticity and of the rotation. Let us indicate with $\eta(t)=\eta_0\sin\omega_{\rm m}t$ with $\eta_0 \ll 1$ the ellipticity introduced by the ellipticity modulator and let us assume that $\eta_0 \gg \Psi_0,\Theta_0$. In the presence of both rotations and ellipticities, the extinguished intensity collected by the photodiode PDE in the ellipticity measurements and rotation measurements is given by
\begin{eqnarray}
&&I_2^{\rm ell}=I_0\left[\sigma^2+|i\eta+i\Psi+\Theta|^2\right]
\nonumber\\
&&I_2^{\rm rot}=I_0\left[\sigma^2+|i\eta+i\Theta+\Psi|^2\right],
\label{eq:ExtinguishedIntensity}
\end{eqnarray}
where $\sigma^2$ is the extinction coefficient of the crossed polarisers and $I_0$ is the intensity transmitted by the cavity, essentially equal to the intensity $I_1$ collected by photodiode PDT. The values of ellipticity or rotation 
are extracted from a Fourier transform of the extinguished intensity. Indicating with $\nu_B$ the rotation frequency of the magnets and with $\nu_{\rm m}$ the modulation frequency of the ellipticity it can be seen, by inspection of Eqs.~(\ref{eq:ExtinguishedIntensity}), that the amplitude of the Fourier components at the frequencies $\nu_{\rm m}\pm2\nu_B$ is linear in the ellipticity or in the rotation, $\Psi_0$ or $\Theta_0$, respectively \cite{Iacopini1979,PVLAS1998}:
\[
I^{\rm ell}_{\nu_{\rm m}\pm2\nu_B}=I_0\,2\,\eta_0\,\Psi_0\qquad{\rm or}\qquad I^{\rm rot}_{\nu_{\rm m}\pm2\nu_B}=I_0\,2\,\eta_0\,\Theta_0.
\]
By demodulating the extinguished intensity at the frequency $\nu_{\rm m}$, the ellipticity and rotation signals can be expressed in terms of the components of the Fourier transform of the demodulated signal as \cite{PVLAS2015}
\begin{equation}
\Psi_0,\Theta_0=\frac{I_{2\nu_B}}{2\sqrt{2\,I_0\,I_{2\nu_{\rm m}}}}.
\label{eq:SolvingLockIn}
\end{equation}

\subsection{Birefringent mirrors}

In previous papers \cite{PVLAS2015,Zavattini2006} it was shown that, as a consequence of the intrinsic birefringence of the cavity mirrors \cite{Brandi1997}, a cross-talk between ellipticity and rotation arises. We are assuming that a laser is frequency-locked to a birefringent cavity and that the input polarisation is aligned with one of the axes of the equivalent wave-plate of the cavity. In the case in which only an ellipticity is generated (pure Cotton-Mouton -- or Voigt -- effect in gases \cite{Rizzo1997}) the electric field component with orthogonal polarisation can be written as
\[
E_{\rm out,2}(\phi)=i\,\Psi_0\,k(\alpha)\left(\frac{E_0T}{1-R}\right)\left(1-i\,\frac{N}{2}\sin\alpha\right)\sin2\phi,
\]
where $T$ is the transmittance of the mirrors, $\alpha$ is the total phase difference (due to the two mirrors) acquired by the two orthogonal polarisation components in a cavity round trip, and
\[
k(\alpha) = \frac{1}{1+N^2\sin^2(\alpha/2)}
\]
is a factor taking into account the fact that, since the extinguished beam experiences a phase shift in the reflection on the mirrors, it cannot be on top of the cavity resonance curve. The electric field has an imaginary component describing the induced ellipticity but also a real component describing a rotation. More in general, in the presence of both an ellipticity $\Psi_0=N\psi_0$ and a rotation $\Theta_0=N\theta_0$ the measured values of $\Psi$ and $\Theta$ are respectively
\begin{equation}
\Psi=k(\alpha)\left[\Psi_0-\frac{N\alpha}{2}\Theta_0\right]\qquad{\rm and}\qquad\Theta=k(\alpha)\left[\Theta_0+\frac{N\alpha}{2}\Psi_0\right].
\label{eq:Signals}
\end{equation}
This means that, if the cavity mirrors are birefringent (and they always are), no meaningful measurement of only ellipticity or rotation can be done. By measuring both, instead, the values of $\Psi_0$ and $\Theta_0$ can in principle be disentangled by solving the two equations (\ref{eq:Signals}) above, provided that $N$ and $\alpha$ are known. These two parameters completely characterise a Fabry-Perot cavity when used as a polarimetric device. In this paper we present a simple method to measure the two parameters. 

A measurement of $N$ is normally obtained by measuring the decay time $\tau_I$ of the intensity transmitted by the cavity, recorded by the photodiode PDT of Fig.~\ref{fig:scheme}:
\begin{equation}
\tau_I=\frac{Nd}{2c},
\label{eq:tau}
\end{equation}
where $d$ is the distance between the cavity mirrors. This measurement usually implies unlocking the laser. In order to measure $\alpha$, one has to resort to experimental configurations in which a birefringence and a dichroism are not simultaneously present. 
In these cases, in fact, only one of the two quantities $\psi_0$ and $\theta_0$ are different from zero. If for example $\theta_0=0$ (pure birefringence, as in gases) one has
\begin{equation}
\Psi_{\rm b}=k(\alpha)\,\Psi_0\qquad{\rm and}\qquad\Theta_{\rm b}=k(\alpha)\,\frac{N\alpha}{2}\,\Psi_0.
\label{eq:CM}
\end{equation}
The ratio of the rotation $\Theta_b$ and of the ellipticity $\Psi_b$ is then
\begin{equation}
R_0=\frac{\Theta_b}{\Psi_b}=\frac{N\alpha}{2},
\label{eq:ratio}
\end{equation}
giving the value of $\alpha$. The same mathematics applies to the complementary case in which only $\theta_0\neq0$ (a pure dichroism -- or, more generally, a rotation):
\begin{equation}
\Theta_{\rm d}=k(\alpha)\,\Theta_0\qquad{\rm and}\qquad\Psi_{\rm d}=k(\alpha)\,\frac{N\alpha}{2}\,\Theta_0.
\label{eq:Faraday}
\end{equation}
The ratio $\Psi_d/\Theta_d$ gives the same value of $R_0$.

\subsection{Frequency dependence}

When doing polarimetry with a Fabry-Perot cavity, since the mirrors in practice always feature some birefringence, one has to pay attention to the different frequency dependences of the two terms appearing in each of the two equations (\ref{eq:Signals}). As the effects are modulated by the rotation of the magnetic field, these equations concern the amplitude of Fourier components of the extinguished intensity, and not dc quantities: the equations (\ref{eq:Signals}) strictly hold only in the limit of zero frequency. 
The first term in each of the two equations (\ref{eq:Signals}) has the same frequency dependence as the electric field transmitted by the cavity, namely that of a first order filter \cite{Uehara1995}. 
The second terms in the two equations, instead, are generated by a second order process. In fact, in order for them to show up, the polarisation component orthogonal to the input polarisation must first be created and only then the effect of the intrinsic birefringence of the mirrors can transform the ellipticity into a rotation and {\em vice versa}. 

\begin{figure}[tbh]
\begin{center}
\resizebox{.8\textwidth}{!}{%
  \includegraphics{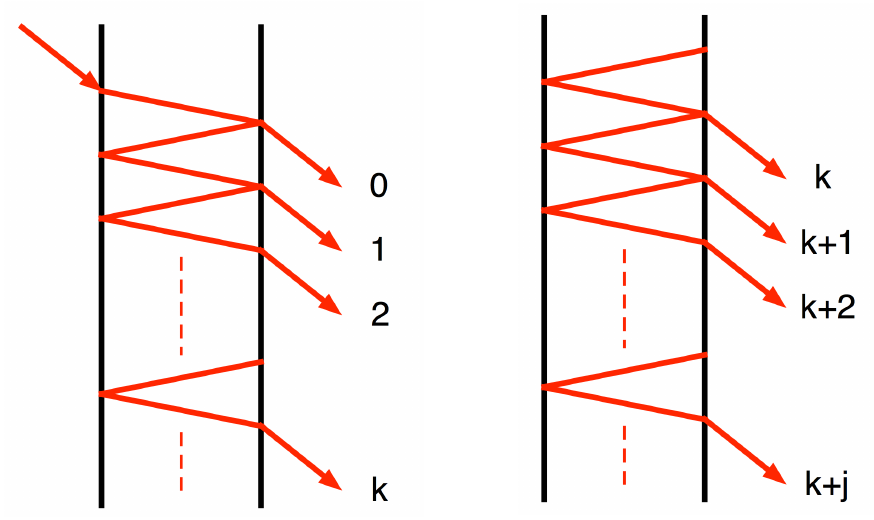}
}\end{center}
\caption{Fabry-Perot cavity used for magnetic polarimetry, schematised as a multiple interference phenomenon. The incidence angle is shown as non-zero only for the sake of the argument; it is brought to zero at the end of the calculation. Left: transmitted beam; Right: extinguished beam.}
\label{fig:interference}
\end{figure}

A more detailed analysis of the phenomenon shows that the non-zero value of $\alpha$ modifies the frequency response of the two signals with respect to the simple first and second order filters. The calculation is performed referring to the usual scheme of multiple interference depicted in Fig.~\ref{fig:interference}, applied to the two cavities, one travelled by light having polarisation equal to the input one, and the other with orthogonal polarisation. The two cavities coincide spatially, but do not interfere. The second cavity has no input beam, and is pumped by the magnetic anisotropy. We analyse the case of the magnetic birefringence that is generated by the rotating magnetic field in gas; the specular case of the dichroism can be treated exactly with the same mathematics. The magnetic birefringence due to the rotating magnets can be schematised as a rotating birefringent wave-plate with a small phase difference $2\psi_0\ll1$ (as we will see, one must have $N\psi_0\ll1$). This corresponds to a maximum ellipticity per passage equal to $\psi_0$. As said above, light in the first cavity is kept on top of the resonance curve by means of a feedback system that locks the laser frequency to the resonance frequency of this cavity. This amounts to say that the multiple beam pattern on the left of Fig.~\ref{fig:interference} is stationary, and not time dependent. In this condition, the electric field at the exit of this cavity is easily calculated as
\[
E_{{\rm out},1}=E_0T\left(1+R+R^2+...+R^k+....\right)=\frac{E_0T}{1-R}=E_0T\,\frac{N}{2},
\]
where $R$ is the reflectance of the mirrors. To calculate the behaviour of the second cavity with orthogonal polarisation, we focus on the $k$-th pair of passages inside the first cavity, with an electric field $\approx E_0\sqrt{T}R^k$. The beam traverses two times the birefringent region, giving a contribution to the electric field inside the second cavity equal, to first order in $\psi_0$, to $2i\psi_0\sin2\phi\,E_0\sqrt{T}R^k$, where $\phi=\omega_B t+\phi_0$ is the time dependent angle between the input polarisation direction and the instantaneous direction of the magnetic field. This contribution exits the second cavity by undergoing multiple reflections, during which a negligible ellipticity is generated back in the first cavity. To synchronise all the partial beams in the output electric field, however, one has to consider contributions that have been generated at previous times separated by the time interval $\tau=2d/c$. The electric field at the exit of the second cavity is then given by
\begin{eqnarray*}
E_{{\rm out},2}(\phi)&\approx&2i\psi_0E_0T\sum_kR^k\sum_jR^je^{-ji\alpha}\sin2\phi_j=\\
&=&E_0\frac{T\psi_0}{1-R}\displaystyle\left[\frac{e^{2i\phi}}{1-Re^{-i\alpha}e^{2i\omega_B\tau}}-\frac{e^{-2i\phi}}{1-Re^{-i\alpha}e^{-2i\omega_B\tau}}\right],
\end{eqnarray*}
where $\phi_j=\phi-j\omega_B\tau$. In this second cavity, the phase factor $e^{-i\alpha}$ is introduced at each round trip to take into account the birefringence of the cavity mirrors: remember that the polarisation is aligned to one of the cavity axes.

Using the Jones matrices, the electric field at the photodiode PDE is obtained as
\[
\mathbf{E}(\phi)=
\left(\begin{array}{cc}0&0\\0&1\end{array}\right)
\cdot
\left(\begin{array}{cc}1&i\,\eta\\i\,\eta&1\end{array}\right)
\cdot
\left(\begin{array}{cc}q&0\\0 &q^*\end{array}\right)
\cdot
\left(\begin{array}{c}E_{{\rm out},1}\\E_{{\rm out},2}\end{array}\right).
\]
In this formula, from left to right, one finds the matrices of the analyser, of the ellipticity modulator, and of the quarter-wave-plate. In this last matrix, $q=1$ for ellipticity measurements, when the wave-plate is out of the optical path and the matrix therefore coincides with the identity matrix, whereas $q=(1+i)/\sqrt{2}$ for rotation measurements when the quarter wave-plate is inserted. The extinguished electric field is then
\[
E_2(\phi)=\frac{E_0T}{1-R}\left[iq\eta+q^*\psi_0\left(\displaystyle\frac{e^{2i\phi}}{1-Re^{-i\alpha}e^{2i\omega_B\tau}}-\frac{e^{-2i\phi}}{1-Re^{-i\alpha}e^{-2i\omega_B\tau}}\right)\right].
\]

By Fourier transforming the intensity recorded by the photodiode PDE demodulated at the frequency $\nu_{\rm m}$ of the ellipticity modulator, and by substituting $2\omega_B$ with a generic signal angular frequency $2\pi\nu$, one is able to obtain the values of the ellipticity $\Psi_b$ and of the rotation $\Theta_b$ by making reference to Eq.~(\ref{eq:SolvingLockIn}). The results of the calculation are expressed as amplitude and phase:
\begin{eqnarray}
&&\displaystyle|\Psi_b|=\frac{T^2}{(1-R)^2}\;\sqrt{\frac{16\,\psi_0^2\,[1-R\cos\alpha(2\cos\delta-R\cos\alpha)]}{\left[1+R^2-2R\cos(\alpha-\delta)\right]\left[1+R^2-2R\cos(\alpha+\delta)\right]}}\nonumber\\
&&\displaystyle\arg(\Psi_b)=\arctan\left[\frac{-R\left[(1+R^2)\cos\alpha-2R\cos\delta\right]\sin\delta}{1+R^2+R\left[R\cos2\alpha-(3+R^2)\cos\alpha\cos\delta+R\cos2\delta\right]}\right]\nonumber\\
&&\displaystyle|\Theta_b|=\frac{T^2}{(1-R)^2}\;\sqrt{\frac{16\,\psi_0^2\,R^2\sin^2\alpha}{\left[1+R^2-2R\cos(\alpha-\delta)\right]\left[1+R^2-2R\cos(\alpha+\delta)\right]}}\nonumber\\
&&\displaystyle\arg(\Theta_b)=\arctan\left[\frac{-2R\cos\alpha+(1+R^2)\cos\delta}{(1-R^2)\,\sin\delta}\right],
\label{eq:FrequencyResponsesB}
\end{eqnarray}
where $\delta=2\pi\nu\tau$. The formulas for the case of the dichroism are obtained from the ones above substituting $\Psi_b$ with $\Theta_d$, $\Theta_b$ with $\Psi_d$ and $\psi_0$ with $\theta_0$:
\begin{eqnarray}
&&\displaystyle|\Theta_d|=\frac{T^2}{(1-R)^2}\;\sqrt{\frac{16\,\theta_0^2\,[1-R\cos\alpha(2\cos\delta-R\cos\alpha)]}{\left[1+R^2-2R\cos(\alpha-\delta)\right]\left[1+R^2-2R\cos(\alpha+\delta)\right]}}\nonumber\\
&&\displaystyle\arg(\Theta_d)=\arctan\left[\frac{-R\left[(1+R^2)\cos\alpha-2R\cos\delta\right]\sin\delta}{1+R^2+R\left[R\cos2\alpha-(3+R^2)\cos\alpha\cos\delta+R\cos2\delta\right]}\right]\nonumber\\
&&\displaystyle|\Psi_d|=\frac{T^2}{(1-R)^2}\;\sqrt{\frac{16\,\theta_0^2\,R^2\sin^2\alpha}{\left[1+R^2-2R\cos(\alpha-\delta)\right]\left[1+R^2-2R\cos(\alpha+\delta)\right]}}\nonumber\\
&&\displaystyle\arg(\Psi_d)=\arctan\left[\frac{-2R\cos\alpha+(1+R^2)\cos\delta}{(1-R^2)\,\sin\delta}\right].
\label{eq:FrequencyResponsesD}
\end{eqnarray}
It is easy to verify that the formulas (\ref{eq:CM}) and (\ref{eq:Faraday}) are obtained from the amplitudes in the limit $\delta\rightarrow0$.

\begin{figure}[tbh]
\begin{center}
\resizebox{.5\textwidth}{!}{%
  \includegraphics{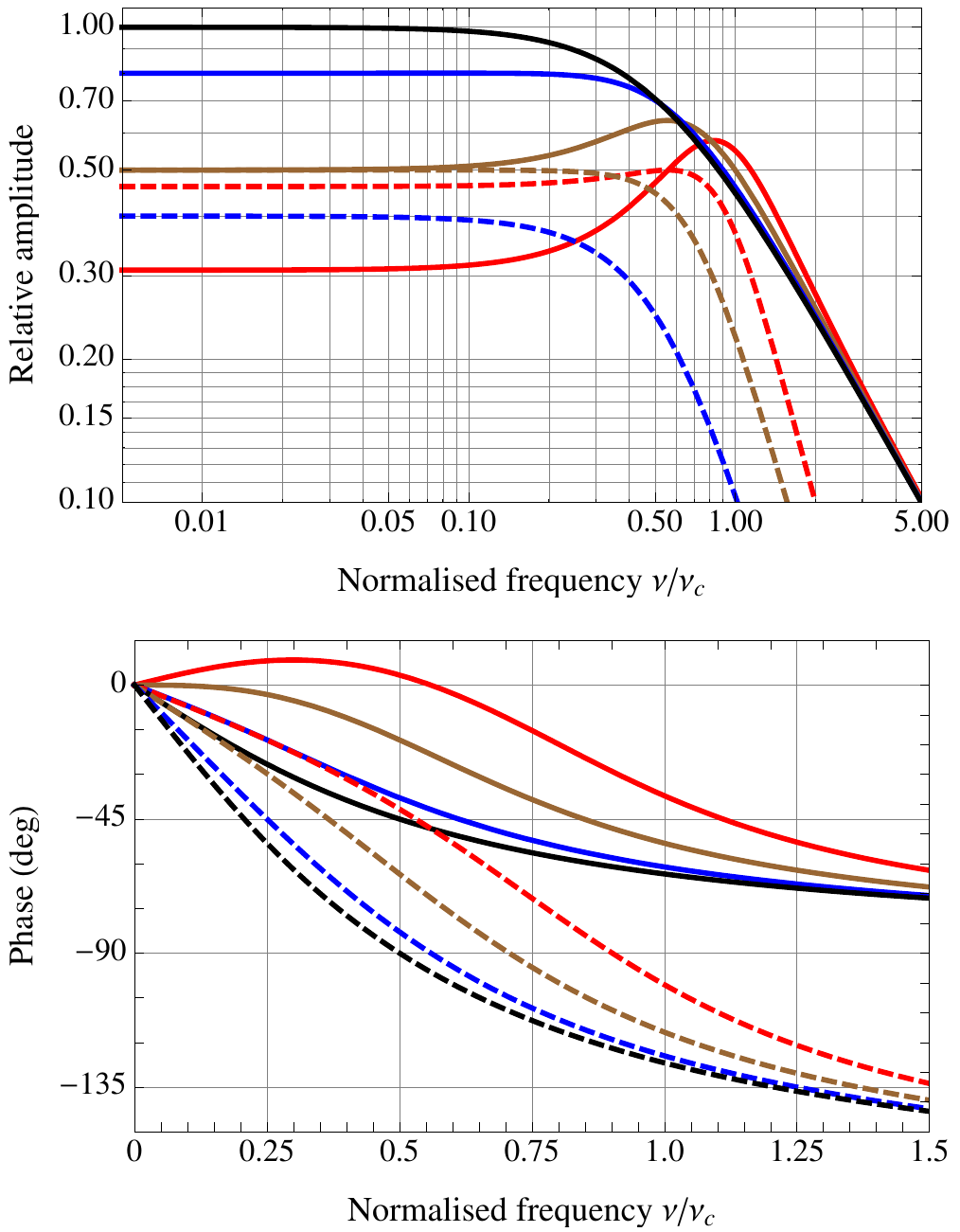}
}\end{center}
\caption{Top panel: Calculated frequency response of the amplitude of the ellipticity (continuous curves) and of the rotation (dashed curves) generated by magnetic birefringence in gas in the PVLAS polarimeter, employing a high-finesse Fabry-Perot cavity with birefringent mirrors.  The frequency scale is expressed in units of the cavity line-width $\nu_c$; the vertical scale is normalised to the low-frequency amplitude of the $R_0=0$ filter [Eq.~(\ref{eq:First})]. The ellipticity curves are drawn for the values of the low-frequency ratio of rotation to ellipticity [Eq.~(\ref{eq:ratio})] $R_0=0$, 0.5, 1.0 and 1.5; the rotation curves have $R_0=0.5$, 1.0 and 1.5. Bottom panel: Calculated frequency response of the phase of the ellipticity (continuous curves) and of the rotation (dashed curves), drawn for the values of $R_0=0$, 0.5, 1.0 and 1.5. The curves have been arbitrarily chosen to start at zero phase and have negative slope.}
\label{fig:FrequencyResponses}
\end{figure}

In Fig.~\ref{fig:FrequencyResponses}, the four equations (\ref{eq:FrequencyResponsesB}) above are plotted as functions of the frequency for various values of $R_0=N\alpha/2$. In the limit $R_0\rightarrow0$, the transfer function of the ellipticity reduces to that of a first order filter \cite{Uehara1995}:
\begin{eqnarray}
\nonumber
\lim_{R_0\rightarrow0}|\Psi_d|&\propto&\frac{T}{\sqrt{1+R^2-2R\cos\delta}}=h_{\rm T}(\nu)\\
\lim_{R_0\rightarrow0}\arg(\Psi_d)&=&\arctan\left[\frac{R\sin\delta}{1-R\cos\delta}\right]=\phi_{\rm T}(\nu).
\label{eq:First}
\end{eqnarray}
In the same limit, the rotation amplitude vanishes; nevertheless, for small $R_0$, the shape of the rotation curve and its phase converge to those of a second order filter:
\begin{equation}
H_{\rm T}(\nu)=h_{\rm T}^2(\nu)\qquad{\rm and}\qquad\Phi_{\rm T}(\nu)=2\phi_{\rm T}(\nu).
\label{eq:Second}
\end{equation}

In this article we present an experimental study of the Fabry-Perot cavity of the PVLAS polarimeter, obtained through the measurements of two magneto-optical effects as a function of frequency. One is the magnetic birefringence in gas. The second is the rotation generated by a Faraday effect in the reflecting layers of the dielectric mirrors of the cavity \cite{Iacopini1983}; for this second measurement, a solenoid coil has been added to the PVLAS apparatus. The choice of the Faraday effect is motivated by the absence of a usable low-energy magnetic dichroism effect in gases. Each of the two sets of data is a complete characterisation of the FP cavity. Both confirm the frequency dependences quoted above.

\section{Experimental set-up and method}

\subsection{The PVLAS polarimeter}

\begin{figure}[tbh]
\begin{center}
\resizebox{0.9\textwidth}{!}{%
  \includegraphics{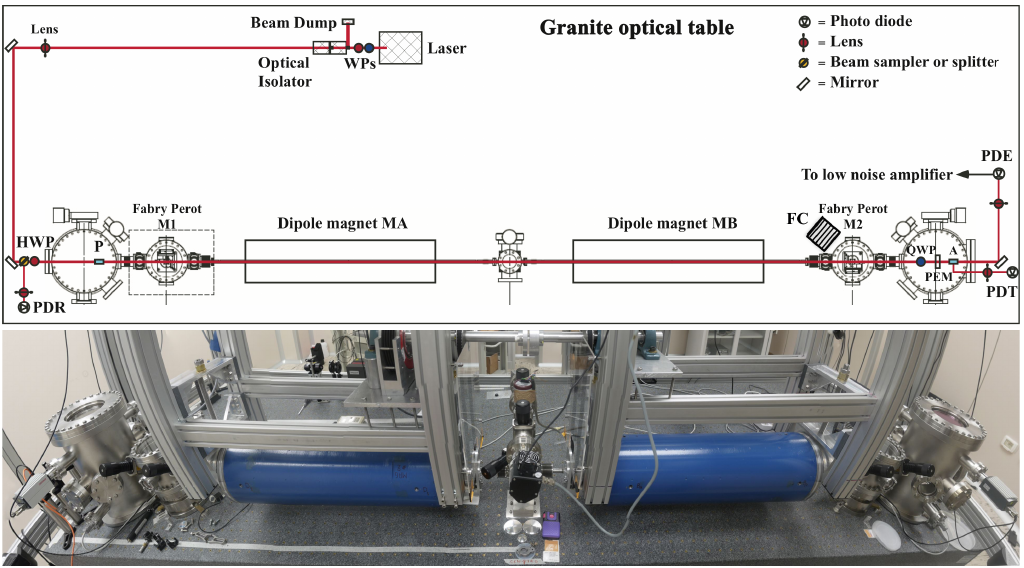}
}
\end{center}
\caption{Top: optical scheme of the apparatus. WPs~=~Wave-plates; HWP~=~Half-wave-plate; PDR~=~Reflection photodiode; P~=~Polariser; Ms~=~Mirrors; QWP~=~Quarter-wave-plate; PEM~=~Photoelastic modulator; A~=~Analyser; PDT~=~Transmission photodiode; PDE~=~Extinction photodiode; FC~=~Faraday coil. Bottom: a picture of the experiment. The two blue cylinders are the dipole magnets.}
\label{fig:experimental}
\end{figure}
\begin{figure}[tbh]
\begin{center}
\resizebox{0.7\textwidth}{!}{%
  \includegraphics{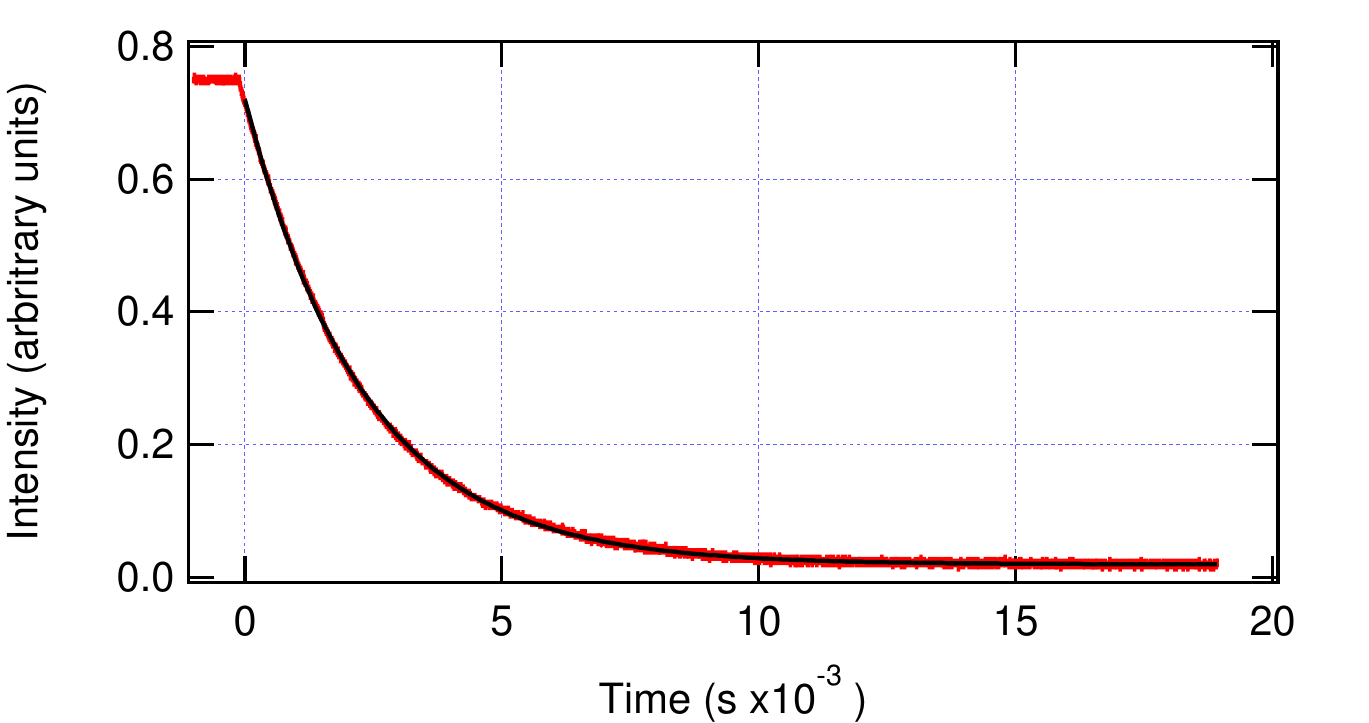}
}\end{center}
\caption{Decay of the intensity $I_1$ transmitted by the Fabry-Perot cavity following switching off of the system that locks the frequencies of the laser and the cavity. The experimental curve is fitted with an exponential function $Ae^{-t/\tau_I}+C$ with $\tau_I=2.32$~ms, corresponding to a finesse ${\cal F}=662\times10^3$.}
\label{fig:decay}
\end{figure}

The set-up of the PVLAS has been presented elsewhere \cite{PVLAS2015}, hence only a brief summary will be given here. A scheme and a picture of the apparatus are shown in Fig.~\ref{fig:experimental}. The polarimeter consists of a 3.3~m long Fabry-Perot cavity inserted between two crossed polarisers. The finesse of the cavity has been measured recording the decay of the intensity $I_1$. The results are reported in Fig.~\ref{fig:decay}. The best fit gives $\tau_I=2.32\pm0.02$~ms, corresponding to a finesse ${\cal F}=(662\pm6)\times10^3$, $N=(421\pm4)\times10^3$ and a line-width $\nu_c=68.0\pm0.6$~Hz. The whole polarimeter, from the polariser to the analyser is kept in high vacuum (pressure better than $10^{-7}$~mbar) or in a pure gas at low pressure. We have verified that the value of the finesse does not change due to the presence of the gas. The light source is a 2~W tuneable laser, frequency-locked to the cavity using the Pound-Drever-Hall method \cite{Cantatore1995}. In the cavity, the beam travels inside glass tubes traversing the bores of two identical dipole permanent magnets (magnetic field lines orthogonal to the light path). The magnetic field inside each 20~mm diameter bore is 2.5~T, each magnetic region having an effective length $L=0.82$~cm long \cite{PVLAS2015}. The magnets rotate around their axes; this modulates both the magnetic birefringence and the dichroism at twice the rotation frequency. Before the analyser, a small ellipticity, modulated at $\nu_{\rm m}=50$~kHz, is added to the beam polarisation by means of a Photo-Elastic Modulator (PEM). A retractable quarter-wave plate is inserted between the FP and the PEM during the rotation measurements. After the analyser, the extraordinary beam and the extinguished one are collected on two photodiodes. The extinguished intensity is demodulated at the frequency of modulation $\nu_{\rm m}$. The signal is then filtered, digitised and Fourier transformed.

\subsection{The measurement method}

In the work presented in this paper, two different measurement configurations have been used. In both configurations, the ellipticity signal $\Psi$ and the rotation signal $\Theta$ have been measured, both in amplitude and phase, as a function of the frequency. In the first configuration, one of the two rotating PVLAS magnets is used to induce a magnetic birefringence in Ar gas at $880~\mu$bar (Cotton-Mouton effect \cite{Rizzo1997}) at frequencies ranging from 1 to 46~Hz (magnet rotation frequencies from 0.5 to 23~Hz). The ellipticity and rotation signals are in this case described by Eqs.~(\ref{eq:FrequencyResponsesB}). In the second configuration, a solenoid coil, positioned outside the vacuum chamber hosting one of the cavity mirrors and roughly aiming at the mirror centre, is used to generate an alternating magnetic field with a significant component orthogonal to the reflecting surface of the mirror, thus generating a Faraday effect \cite{Iacopini1983}. We analyse the data in the same frequency range as above. With a 1~A current, the coil produces a magnetic field $\approx1$~G at a distance of 15~cm from the mouth of the coil (distance to the mirror). Precise values of the magnitude and of the orientation of the magnetic field at the position where the light beam impinges on the mirror are unknown, but on the other hand unnecessary. Since in this second experiment the vacuum vessel is kept in vacuum, no gas birefringence is generated, and the measured rotation and ellipticity signals are given by Eqs.~(\ref{eq:FrequencyResponsesD}). In the case of the Cotton-Mouton effect, rotating the magnet at a frequency $\nu_B$ generates signals at frequency $2\nu_B$. In the case of the Faraday effect, a magnetic field oscillating at a frequency $\nu_F$ generates signals at the same frequency. In this case, the Frequency Response function of an Agilent 35670A Dynamic Signal Analyzer has been employed. The frequency response is obtained as the coherent average of fast 0.16~s/Hz sweeps over 400 points and 50~Hz frequency range. In the case of the Cotton-Mouton effect, the phases of the signals are measured with respect to a trigger signal generated by a contrast sensitive diode in correspondence of the passage of a mark drawn on the external surface of the rotating magnet. The diode has a response time $<50~\mu$s, corresponding to a maximum phase uncertainty $<0.8^\circ$ at 46~Hz. Note that the resulting phases depend on the angle between the polarisation direction and the diode position. In the case of the Faraday effect, the phases are measured with respect to the current flowing through the coil. In both cases, a small correction has been subtracted from the measured phase, due to the frequency response of the lock-in amplifier used to demodulate the signal from the diode PDE collecting the extinguished intensity. The amplitude of the signals measured in the Faraday effect has been normalised to the intensity of the current in the coil.

\section{Results and discussion}

\subsection{Magnetic birefringence in gas}

\begin{figure}[tbh]
\begin{center}
\resizebox{0.8\textwidth}{!}{%
  \includegraphics{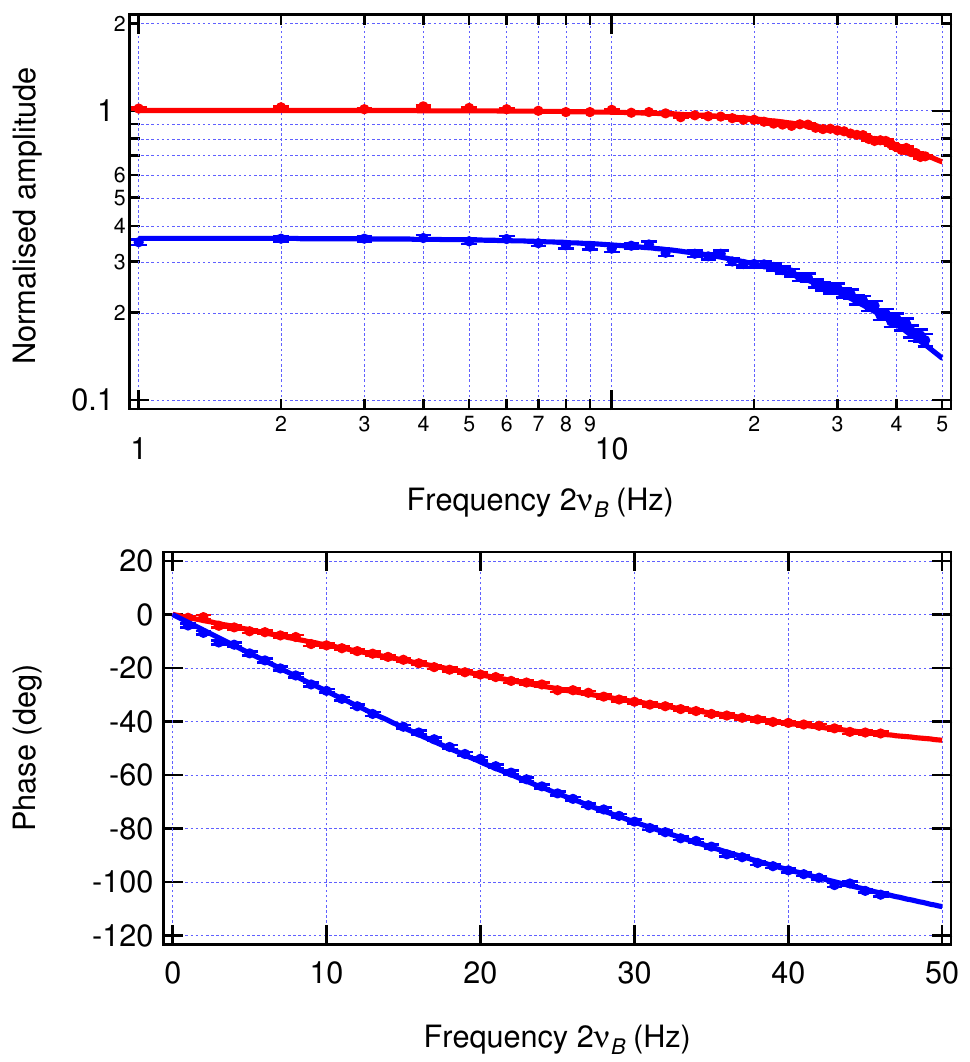}
}
\end{center}
\caption{Top panel: relative amplitude of the ellipticity signal $\Psi_b$ (upper curve) and of the rotation signal $\Theta_b$ (lower curve) measured as a function of frequency for the Cotton-Mouton effect of $880~\mu$bar of Ar gas. Bottom panel: phases of the ellipticity signal $\Psi_b$ (upper curve) and of the rotation signal $\Theta_b$ (lower curve). The continuous lines are the global fit obtained with Eqs.~(\ref{eq:FrequencyResponsesB}). The values of the finesse ${\cal F}$ and of the phase difference $\alpha$ are obtained as ${\cal F}=640\times10^3$ and $\alpha=1.78~\mu$rad.}
\label{fig:CottonMouton}
\end{figure}

In Fig.~\ref{fig:CottonMouton}, the relative amplitude and phase of the ellipticity and rotation signals measured as a function of the frequency in the Cotton-Mouton effect of $880~\mu$bar of Ar gas are shown. Integration time was 256~s/point for both ellipticity and rotation measurements. A constant phase, measuring the zero-frequency relative position of the signals and the trigger, has been subtracted from the phase data, so as to have both curves starting at zero phase. The experimental points are fitted simultaneously with the four functions of Eqs.~(\ref{eq:FrequencyResponsesB}). The fit provides a unique value for the reflectance $R$ and hence for the finesse ${\cal F}$ [see Eq.~(\ref{eq:finesse})], and for the phase difference $\alpha$ appearing in the expressions of the four curves. The values found are ${\cal F}=(640\pm4)\times10^3$ and $\alpha=(1.78\pm0.02)~\mu$rad, with a normalised $\chi^2_{\rm o.d.f.}=181/174$. The value obtained for the finesse is about 4\% smaller than the value obtained from the analysis of the decay of the intensity transmitted by the FP cavity (see Fig.~\ref{fig:decay}), and is compatible with that within three standard deviations. The uncertainties used in the fit are the piecewise standard deviations of the residuals obtained by fitting the four curves separately. In a first tentative of a global fit, the residuals of the phase data exhibited a marked linear behaviour of a few degrees over the whole frequency interval. This behaviour can be attributed to the fact that, during the measurements, the polarisation direction of the light entering the Fabry-Perot cavity is varied by small quantities to compensate for the slow drift of the static birefringence of the cavity. We have then added two linear functions to the two phase fit functions. The values of the slopes obtained through the fit are $(0.1^\circ\pm0.01^\circ)$~Hz$^{-1}$ for the phase of the ellipticity, and $(0.05^\circ\pm0.01^\circ)$~Hz$^{-1}$ for the phase of the rotation. Note that the duration of the ellipticity and rotation measurements were, respectively, eight hours and four hours, leading to an identical drift of $160~\mu$deg/s in the two measurements. This strongly supports the suggested interpretation. It is worth noting that the value of $\alpha$ is small enough that fitting simultaneously the four data sets with the expressions of the first and second order filters (\ref{eq:First}) and (\ref{eq:Second}) still produces a reasonable fit, with a similar $\chi^2$ probability, but at the expenses of an unreasonable 20\% reduction of the value of ${\cal F}$ and of completely incompatible drifts of the ellipticity and rotation phases. In this case, $\alpha$ can be extracted from the low frequency ratio $R_0$ of the amplitudes of rotation and ellipticity [see Eq.~(\ref{eq:ratio})].

\subsection{Magnetic rotation}

\begin{figure}[tbh]
\begin{center}
\resizebox{0.8\textwidth}{!}{%
  \includegraphics{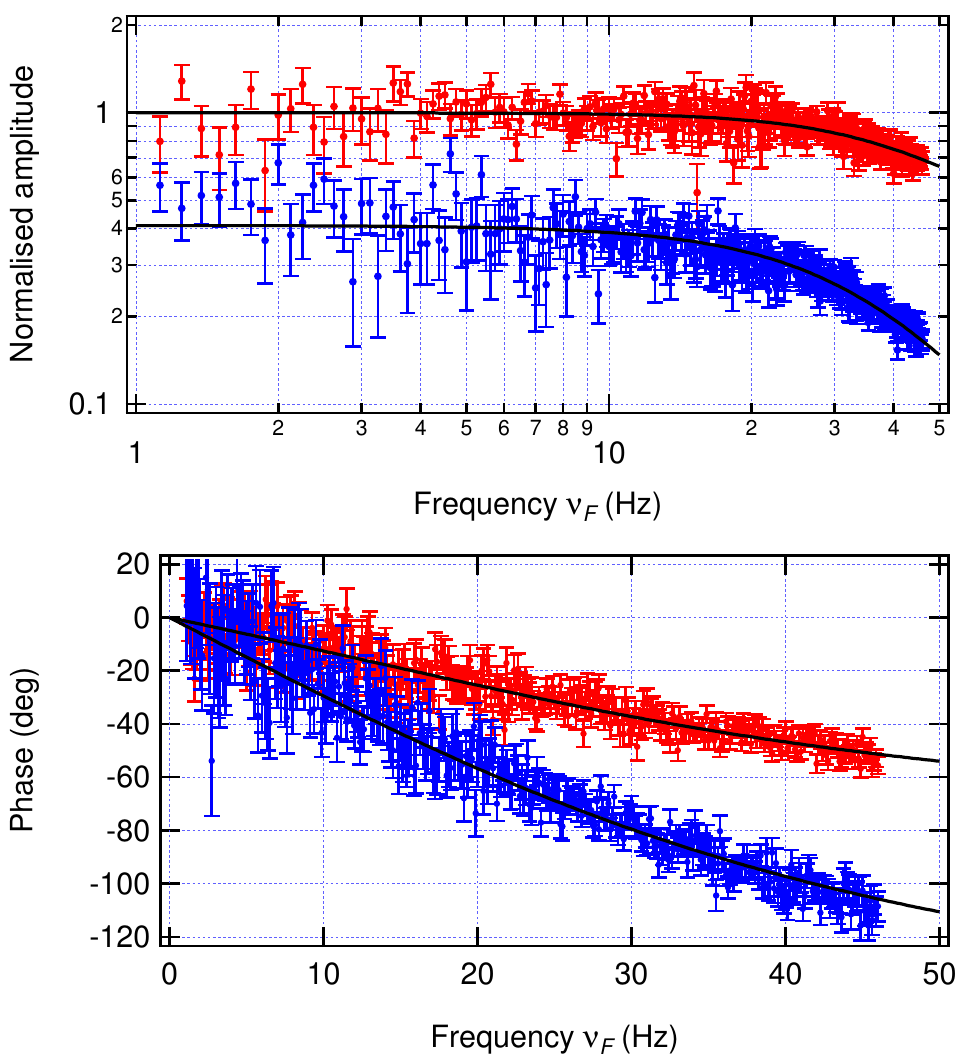}
}
\end{center}
\caption{Top panel: relative amplitude of the rotation signal $\Theta_d$ (upper curve) and of the ellipticity signal $\Psi_d$ (lower curve) measured as a function of frequency for the Faraday effect on the reflecting surface of a mirror of the FP cavity. Bottom panel: phases of the rotation signal $\Theta_d$ (upper curve) and of the ellipticity signal $\Psi_d$ (lower curve). The continuous lines are the global fit obtained with Eqs.~(\ref{eq:FrequencyResponsesD}). The values of the finesse ${\cal F}$ and of the phase difference $\alpha$ are obtained as ${\cal F}=691\times10^3$ and $\alpha=1.87~\mu$rad.}
\label{fig:Faraday}
\end{figure}

In Fig.~\ref{fig:Faraday}, the relative amplitude and phase of the rotation and ellipticity signals measured as a function of the frequency $\nu_{\rm F}$ of the current in the Faraday coil are shown. Integration time was 18~s/point for the rotations measurement, and 50~s/point for the ellipticity measurement. In this second experiment, an alternating current with amplitude $\approx1$~A and frequency $\nu_F$ sweeping between 0 and 50~Hz is fed to the coil FC of Fig.~\ref{fig:experimental}. The amplitude data have been normalised to the current through the coil. A constant phase, measuring the zero-frequency relative position of the signals and the trigger, has been subtracted from the phase data, so as to have both curves starting at zero phase. The experimental points are fitted with the four functions of Eqs.~(\ref{eq:FrequencyResponsesD}). The fit provides a unique value for the finesse ${\cal F}$ and the phase difference $\alpha$ appearing in the expressions of the four curves. The values found are ${\cal F}=(691\pm8)\times10^3$ and $\alpha=(1.87\pm0.02)~\mu$rad with a normalised $\chi^2_{\rm o.d.f.}=1472/1434$. The value obtained for the finesse is about 4\% larger than the value obtained from the analysis of the decay of the intensity transmitted by the FP cavity (see Fig.~\ref{fig:decay}), and is compatible with that within three standard deviations. The value of $\alpha$ is 5\% larger than the one found in the Cotton-Mouton experiment described above; the two values are compatible within three standard deviations. This small difference could be accounted for by the fact that the two datasets were taken in different days and that we know that $\alpha$ is subject to small drifts. As in the case of the Cotton-Mouton measurement, the uncertainties used in the fit are the piecewise standard deviations of the residuals obtained by fitting the four curves separately. Differently from the Cotton-Mouton case, no linear addition to the phase fit function was necessary. This is consistent with the interpretation of the feature observed in the Cotton-Mouton effect: in fact, in the case of the Faraday measurements, the phase is electronically defined. 
By fitting the four curves with the expressions of the first and second order filters (\ref{eq:First}) and (\ref{eq:Second}) we obtained ${\cal F}=594\times10^3$, with a $\chi^2$ probability of $5\times10^{-3}$, justifying the necessity of introducing the parameter $\alpha$.

\section{Conclusions}

In this paper we have presented two experimental characterisations of an optical Fabry-Perot cavity used for polarimetry. The studies provide values for the finesse of the cavity and for the phase difference acquired by light upon reflection on the dielectric mirrors of the cavity. The measurements are performed while the laser is  frequency-locked to the cavity. The data are analysed in terms of the frequency response of the ellipticity and rotation signals, whose theory is developed and presented. The theoretical frequency responses show a marked deformation from the expressions of the first and second order filters, reducing to these last expressions for $R_0=\frac{N\alpha}{2}\ll1$.

It should be noted that the Cotton-Mouton characterisation of the cavity requires introducing a gas in the Fabry-Perot enclosure whereas the Faraday characterisation is less intrusive as it requires no interventions on the Fabry-Perot. The Faraday effect on the mirrors can therefore be used to monitor $R_0$ during polarimetric measurements.

\section{Acknowledgments}
This work has been performed as a development activity of the PVLAS experiment intended to optimise the performances of the apparatus built to measure for the first time the quantum magneto-optical properties of vacuum. We warmly thank our colleagues of the PVLAS collaboration U. Gastaldi, E. Milotti, R. Pengo and G. Ruoso for helpful discussions and encouragement and for critical reading of the manuscript.

%
%


\begin{thebibliography}{99}
\bibitem{PVLAS2015}
F. Della Valle {\em et al.} (PVLAS collaboration), Eur. Phys. J. C {\bf76}, (2016) 24.
\bibitem{Q&A2007}
2.	S.-J. Chen, H.-H. Mei and W.-T. Ni (Q\&A collaboration), Mod. Phys. Lett. A {\bf22}, (2007) 2815.
\bibitem{Cadne2014}
A. Cad\`ene, {\em et al.} (BMV collaboration), Eur. Phys. J. D {\bf68}, (2014) 16.
\bibitem{Fan2017}
X. Fan {\em et al.} (OVAL collaboration), arXiv:1705.00495v1.
\bibitem{Uehara1995}
N. Uehara and K. Ueda, Appl. Phys. B {\bf61}, (1995) 9.
\bibitem{Berceau2010}
R. Berceau {\em et al.}, Appl. Phys. B {\bf100}, (2010) 803.
\bibitem{Zavattini2006}
G. Zavattini {\em et al.}, Appl. Phys. B {\bf83}, (2006) 571.
\bibitem{Iacopini1979}
E. Iacopini and E. Zavattini, Phys. Lett. {\bf85B}, (1979) 151.
\bibitem{Iacopini1981}
E. Iacopini {\em et al.}, Nuovo Cimento {\bf61 B}, (1981) 21.
\bibitem{BFRT1993}
R. Cameron {\em et al.} (BFRT collaboration), Phys. Rev. D {\bf47}, (1993) 3707.
\bibitem{PVLAS1998}
D. Bakalov {\em et al.} (PVLAS collaboration), Quantum Semiclass. Opt. {\bf10}, (1998) 239.
\bibitem{PVLAS2008}
M. Bregant {\em et al.} (PVLAS collaboration), Phys. Rev. D {\bf78}, (2008) 032006.
\bibitem{PVLAS2013}
F. Della Valle {\em et al.} (PVLAS collaboration), New J. Phys. {\bf15}, (2013) 053026.
\bibitem{PVLAS2014}
F. Della Valle {\em et al.} (PVLAS collaboration), Phys. Rev. D {\bf90}, (2014) 092003.
\bibitem{Vallet1999}
For a review of the Malus interferometer see M. Vallet {\em et al.}, Opt. Commun. {\bf168}, (1999) 423.
\bibitem{Jones1948}
R.C. Jones, J. Opt. Soc. Am. {\bf38}, (1948) 671, and references therein.
\bibitem{BornWolf}
M. Born and E. Wolf, \textit{Principles of Optics} 6th ed. (Pergamon Press, Oxford 1989).
\bibitem{DellaValle2014OE}
F. Della Valle {\em et al.}, Opt. Express {\bf22}, (2014) 11570.
\bibitem{Brandi1997}
F. Brandi {\em et al.}, Appl. Phys. B {\bf65}, (1997) 351.
\bibitem{Rizzo1997}
For a review, see C. Rizzo, A. Rizzo and D.M. Bishop, Int. Rev. Phys. Chem. {\bf16}, (1997) 81.
\bibitem{Iacopini1983}
E. Iacopini, G. Stefanini and E. Zavattini, Appl. Phys. A {\bf32}, (1983) 63.
\bibitem{Cantatore1995}
G. Cantatore {\em et al.}, Rev. Sci. Instr. {\bf66}, (1995) 2785, and references therein.
\end{thebibliography}
\end{document}